\documentclass{article}

\begin{document}
\renewcommand{\baselinestretch}{2}
 \parskip .2in

\newcommand{\e}{{\rm e}}

\title{On the Action, Topology and Geometric Invariants in Quantum Gravity}
\author{\sc A. Coley {\small\em Dept. Mathematics, Dalhousie University, B3H 3J5 Halifax, Canada}}

\maketitle

\begin{abstract}

The action in general relativity (GR), which is an integral over the manifold plus an integral over the boundary,
is a global object and is only 
well defined when the
topology is fixed. Therefore,
to use the action in GR and in most approaches to quantum gravity (QG)
based on a covariant Lorentzian  action,
there needs to exist a prefered (global) 
timelike vector, and hence 
a global topology $R \times S^3$, for it to make sense.
This is especially true
in the Hamiltonian formulation of QG.
Therefore, in order to do canonical quantization,
we need to know the topology,
appropriate boundary conditions and
(in an open manifold) the conditions at infinity,
which affects the fundamental geometrical scalar invariants of the spacetime
(and especially those which may occur in the QG action).

\end{abstract}



\newpage

{\em{Introduction}}:
In Einstein's general relativity (GR) matter and geometry are
intimately linked, replacing the previous Newtonian paradigm of matter dynamics occuring on a fixed background. 
However, unlike in GR,  time and space are 
not on an equal footing in quantum mechanics (time is treated classically 
whereas space is associated with a quantum description).
It is questionable as to whether modern theories of quantum gravity  (QG) respect Einstein's
revolutionary way of interpreting  gravitational physics geometrically \cite{QGrefs}.

Most approaches to QG are based on a covariant Lorentzian  action.
Unlike most field theories, in GR the situation is complicated
by the fact that the Einstein-Hilbert
action includes a surface term. In most derivations
of the gravitational Hamiltonian, the surface term is ignored. This results
in a Hamiltonian which is just a multiple of a constraint. One must then add
to this constraint appropriate
surface terms so that its variation is well defined \cite{HawkingHorowitz},
and the action  yields the correct equations of motion subject only to the
condition that the induced three metric and matter fields
on the boundary are held fixed.
The action 
is well defined for spatially compact geometries, but diverges
for noncompact ones.

{\em{Action:}}
The dynamical content of GR is fully expressed by Einstein's FE (EFE).  Nevertheless, even in a purely classical (i.e., 
non-quantum) context, it is convenient and useful for many purposes to have 
Lagrangian and Hamiltonian formulations of GR.  The 
interpretation that EFE describes the 
evolution of a `spatial metric', with ``time'' is perhaps best expressed via 
the Hamiltonian formulation. Moreover, most prescriptions for formulating a quantum field theory 
of gravitation require that the  associated classical theory be expressed in a 
Lagrangian or Hamiltonian form.

For GR, the field variable is the spacetime metric, $g_{ab}$, defined on a 
four-dimensional (4D) manifold, $M$.  In this case, the natural volume element in the integrals 
is the volume element which  itself depends on the field variables $g_{ab}$,
and hence its variation must be 
taken into account when calculating functional derivatives.
In the canonical approach  \cite{Wald} a family of spacelike surfaces is introduced and used to 
construct a Hamiltonian and canonical equal-time commutation relations
(which is appropriate for strong gravitational
fields and is supposed to ensure unitarity).  However, the split into three spatial dimensions 
and one time dimension seems to be contrary to the whole spirit of GR \cite{HawkingIsr}.  
Moreover, it restricts the topology of spacetime to be the product of the 
real line with some 3D manifold, whereas one would 
expect that QG would allow all possible topologies of
spacetime including those which are not products.  It is precisely these 
other topologies that seem to give the most interesting effects.
It is usually implicitly assumed either that the 3-geometry is compact (a `closed' universe) or
that the gravitational and matter fields die off in some suitable way at
spatial infinity (the asymptotically flat space) {\cite{F2}.
Under variations of the metric which vanish and whose normal derivaties also vanish on
$\partial M$, the boundary of a compact region $M$, this action is stationary if and
only if the metric satisfies the EFE.

However this action is not an extremum if one allows
variations of the metric which vanish on the boundary but whose normal 
derivatives do not vanish there \cite{HawkingIsr}. The reason is that the Ricci curvature scalar $R$
contains terms which are linear in the second derivatives of the metric.  By
integration by parts, the variation in these terms can be converted into an
integral over the boundary which involves the normal derivatives of the 
variation on the boundary.  In order to cancel out this surface integral, and 
so obtain an action which is stationary for solutions of the EFE 
under all variations of the metric that vanish on the boundary,
one has to add to the action a term of the form \cite{HawkingIsr}:

\begin{eqnarray}
I & = &\frac{1}{16 \pi G} \int_M (R-2 \Lambda)(-g)^{1/2} \,\, d^4 x+ 
\int_M L_m(-g)^{1/2} \,\, d^4 x \nonumber \\
 & & + \frac{1}{8 \pi G} \int_{\partial M} K(\pm h)^{1/2} \,\, d^3 x + C, 
\end{eqnarray}
where $K$ is the trace of the second fundamental form of the boundary, $h$ is
the induced metric on the boundary, the plus or minus signs are chosen 
according to whether the boundary is spacelike or timelike, and $C$ is a 
term which depends only on the boundary metric $h$ and not on the values of 
$g$ at the interior points (which must be considered in open
universes, including asymptotically flat space-times, but may be set to
zero in closed universes).
 
The action above is a global object and is
well-defined only if the global topology is fixed as $R \times S^3$.
In a sense the Lagrangian formulation of a field theory is ``spacetime covariant''. On the other hand, a Hamiltonian formulation of a field
theory necessarily requires a global breakup of the spacetime into space and time.

For closed universes
we only have the two volume integrals in the above, and these terms
would be finite even if $M$ were chosen to be the entire spacetime (for
suitable choices of the matter Lagrangian $L_m$).  In the open universe case the
boundary terms enter the theory in a fundamental way, and there is no good way 
to decide what these terms should be for arbitrary open universes. In addition, the 
surface integral above is also
more complicated in open universes, since in this case it must contain timelike
portions \cite{F3}.

We note that the  boundary term $\partial M$ is added precisely to cancel the surface terms
and exactly produce the  EFE of GR. Therefore, in GR, the EFE
are more fundamental than the action.
This is of importance
mathematically, since the
EFE and the EL equations derived from action can differ, 
different actions can give rise to the same FE, and
boundary terms can affects geodesics \cite{F4,F5}.

{\em{Topology:}}
Therefore, to know the local evolution (Euler-Lagrange (EL)  field equations (FE)), we need to know the topology,
appropriate boundary conditions and
(in an open manifold) the conditions at infinity. 
A global topology $R \times S^3$  is considered, where  $S^3$ is usually assumed compact. 
Regarding the boundary conditions, 
the action certainly
makes more sense in a closed universe.
The surface integral is more complicated in open universes, in which boundary terms enter in a more fundamental 
way (and it is not known in general what these terms should be). 
Therefore, there are problems with boundary conditions at infinity
for an open manifold: we need to know where infinity is (definition), and conditions at infinity (which might be
timelike or null). 
In an open or closed universe we need to add surface terms on a case-by-case basis 
(e.g. different for each type of spacetime). None of this is
really acceptable in GR. 
In particular, these issues lead to philosophical problems in cosmology. \cite{F1}

Clearly, in order to do canonical quantization additional spacetime structure is needed.
In GR,  there is no background
geometry.  The space-time metric itself is the fundamental
dynamical variable. The canonical and the covariant approaches have adopted
dramatically different attitudes to face these problems. 
In many {\em{covariant}} approaches to QG (with fields evolving on a fixed background \cite{F0}), such as string theory \cite{GQ}, the spacetime metric
is split into a kinematical background and dynamical fluctuations.
The fundamental degrees of freedom and the short scale dynamics
in the final quantum theory are quite different to those of GR and 
classical spacetimes and  gravitons only emerge in a suitable limit
(although the resulting quantum GR turns out to
be non-renormalizable).
However, background independence is important in modern approaches to QG \cite{LQG},
in which the emphasis is on
preserving the geometrical character of GR.

In the {\em{canonical}} approach 
the Hamiltonian formulation of
GR is used and a
fixed (`spatial') three-manifold (usually assumed compact). 
The very first step of the canonical quantization program requires a
splitting of space-time into space and time thereby, as noted earlier, doing grave
injustice to space-time covariance that underlies general
GR. [Therefore, we only attempt to quantize a subset
of spacetimes (e.g., with a global topology $R \times S^3$)]. 
This is a valid concern, but 
successful background independent approaches to
quantum GR (such as LQG and causal dynamical triangulation)
accepts this price.

We also note that {\em{analytic continuation}} (a generalization of a Wick rotation) is often
used in computions in quantum theories.
There is clearly a restriction on
the class of  Lorentzian metrics   in QG, 
and hence on the (real) gravitational degrees of freedom,  
by assuming the existence of such a Wick-rotation.
It is expected that a
Lorentzian spacetime that allows for such an analytic continuation is necessarily  globally 1+3 (and hence
I-non-degenerate) \cite{F6}. In spacetimes with a topology $R \times S^3$
there is an absolute time function, since
the spacetime admits a foliation by globally space-like hypersurfaces,
which restricts the possible existence of closed time-like curves; this is not 
necessarily the case in the
supersymmetric Godel solutions in 
string theory \cite{Gauntlett}.

{\em{Geometric Invariants:}}
Therefore, in the canonical approach to QG there exists a unique time (and space and time are
essentially treated independently; therefore the structure of the Lorentzian
manifold is not fully utilized). A Lorentzian spacetime with global topology $R \times S^3$ is  
$\mathcal{I}$-non-degenerate and thus completely classified
by its set of scalar polynomial curvature invariants \cite{invariants}. 
In this case all gravitational degrees of freedom are curvature invariants.
For example, in many theories of fundamental physics there are
geometric classical corrections to GR. Different polynomial
curvature invariants (constructed from the Riemann tensor and its covariant derivatives) are required to compute different loop-orders of
renormalization of the Einstein-Hilbert action. In specific quantum
models such as supergravity there are particular allowed local counterterms \cite{GQ}.

It was proven  (in 4D, and it is also likely true in arbitrary dimenions)
\cite{invariants} that a spacetime
metric is either $\mathcal{I}$-non-degenerate or a degenerate Kundt metric.
A Lorentzian degenerate Kundt spacetime \cite{kundt} 
is not completely classified
by its set of scalar polynomial curvature invariants \cite{invariants}.
The higher-dimensional Kundt class of
spacetimes  \cite{kundt} are genuinely Lorentzian and have many
mathematical properties quite different from their Riemannian counterparts.
For example, they have important
geometrical information that is not contained in the scalar invariants and, in principle,
the Einstein-Hilbert action may require geometric corrections 
that are not scalar invariants.
In particular, the physical fundamental
properties that do not depend only on scalar invariants may lead
to  interesting and novel physics in models of QG and particularly in string
theory.

A Lorentzian manifold admitting an indecomposable but non-irreducible
holonomy representation (i.e., with a one-dimensional invariant lightlike
subspace) is a degenerate Kundt (degenerately reducible) spacetime, which contains
the VSI and (non locally homogeneous) CSI subclasses (in which all of the scalar invariants are
zero or constant, respectively) as special cases. It is perhaps within string
theory that the full richness of Lorentzian geometry is realised,
where the Kundt spacetimes (which include, for example, VSI generalized pp-wave spacetimes
and CSI generalized $AdS_5\times S^5$ spacetimes, which are not globally $R \times M$)
may play a fundamental role. 
Solutions of the classical FE for which the counter terms required to regularize quantum fluctuations vanish
are of importance because they offer insights into the behaviour of the full quantum theory of gravity
(regardless
of what the exact form of this theory might be).
A classical metric is called universal if the quantum
correction is a multiple of the metric \cite{universal}, and consequently 
such metrics can be interpreted as
having vanishing quantum corrections to all loop orders and are automatically solutions
to the quantum theory. 
In particular, VSI and CSI  spacetimes are exact solutions
in string theory to all perturbative orders in the string tension scale \cite{PRL}.

Finally, it should be noted that many non-fundamental theories are not derived from an action. Hamiltonian systems are non-dissipative. 
For example, the Navier-Stokes 
fluid equations and statistical mechanics are
obtained by coarse graining.
An action principle may not be necessary
(or even possible) for QG or at least a low energy effective version of the theory.
The results of averaging the geometry, however, are expected to be far from trivial, since the
EFE are highly non-linear.
The averaging problem in GR has been studied in \cite{F7}. So in the same sense that
(linear) QED is averaged   to obtain Maxwell eqns, QG should be averaged to get GR
{\cite{F8}. This leads to the problem of 
averaging  at the level of the action \cite{F9}.

\newpage


}

\begin{thebibliography}{99}
\bibitem{F1} In particular, these issues lead to philosophical problems in cosmology.
In an open universe it is necessary to specify condition at infinity (outside of our particle horizon), which is absurd. 
[In addition, how do
we choose when and where to set boundary conditions and initial conditions? 
If inflation  occurs, the Universe is approximately spatially homogeneous at large distance and we 
then face additional problems \cite{MacTaub}].
\bibitem{F2}  
Teitelbaum \cite{Teitel}
treated the boundary conditions when $M$  is not compact. 
The boundary integrals are required to vanish, but the necessary additional
surface terms may only cancel for asymptotically flat spacetimes.  
In particular, the appropriate fall off rate is not fast enough to allow integration
by parts for 
cosmological models.
\bibitem{F5}
In particular,
the FE of a theory and the action principle approach are not mathematically 
equivalent. The
action principle is usually
only used as a computational tool to infer the evolution FE. 
Historically physicists (who developed 
action principles in 
mechanics and general physics) were  motivated by teleology
(since the motion of the physical system is determined in the 
action principle formulation 
by the knowledge of both the initial and
final states of the system)  \cite{BT}.
\bibitem{F7}
In cosmology it is assumed that the real complex lumpy universe with a discrete
matter distribution can be
adequately approximated by an ``effective averaged out'' stress-energy
tensor. 
The averaging problem in cosmology is of considerable importance for the correct
interpretation of cosmological data. A rigorous mathematical definition of averaging in
a cosmological model is necessary. In general, a cosmological spacetime is completely characterized
by its scalar curvature invariants, and a particular spacetime averaging
scheme based entirely in terms of scalar curvature
invariants was proposed in \cite{ColeyAv}.
\bibitem{F9}
An averaged Lagrangian approach (variational principle) for the derivation of Isaacson's
equation (for metrics describing (weak)
gravitational radiation, by treating the gravitational perturbations
using the WKB version of the high-frequency approximation), 
by applying the averaged Lagrangian method of Whitham, 
was formulated in \cite{MacTaub2}. 
It was 
shown that an averaged high-frequency Lagrangian of
a similar form  can be derived by applying
the  macroscopic gravity averaging scheme  \cite{Roustam}. 
That is, by applying the space-time averaging
procedure and the rules of splitting out the products of geometric
objects developed within macroscopic gravity theory \cite{Averaging}
to the
(microscopic) action of GR, an
averaged (macroscopic) action which upon the variation of the macroscopic
metric yields the
FE of macroscopic gravity was obtained. 




----------------------------
\end{thebibliography}
\end{document}